\documentclass{article}

\newcommand{\ie}{\mbox{i.\,e.\,\ }}
\newcommand{\iec}{\mbox{i.\,e.\,}}

\newcommand{\eg}{\mbox{e.\,g.\,\ }}
\newcommand{\egc}{\mbox{e.\,g.\,}}
\newcommand{\etc}{etc.\,\ }

\newcommand{\vctr}[1]{\ensuremath{\mathbf{ #1 }}}

\newcommand{\dr}[1]{\ensuremath{\mathrm{d} #1\,}}

\newcommand{\pbp}[2]{\ensuremath{\frac{\partial #1}{\partial #2}}}

\newcommand{\ket}[1]{\ensuremath{\left|  #1 \right\rangle}}

\newcommand{\bk}[2]{\ensuremath{\left\langle #1 | #2 \right\rangle}}

\newcommand{\tpk}[2]{\ensuremath{\ket{#1}\!\otimes\!\ket{#2}}}

\newcommand{\hilbert}[1]{\ensuremath{\mathcal{#1}}}

\newcommand{\comm}[2]{\ensuremath{\left[ #1 , #2 \right]}}

\newcommand{\lgr}{\ensuremath{\mathcal{L}}}
\newcommand{\gpt}{\ensuremath{\mathcal{A}}}
\newcommand{\gfs}{\ensuremath{\mathcal{F}}}

\begin{document}
\title{The Quantization of Gravity --- an introduction}

\author{David Wallace\thanks{email address: david.wallace@merton.ox.ac.uk} \\ Centre for Quantum Computation \\ The Clarendon Laboratory, Oxford University \\
Parks Road, Oxford OX1 3PU, UK}
\date{April 3, 2000}

\maketitle

\begin{abstract}
This is an introduction to quantum gravity, aimed at a fairly general audience and concentrating on what have historically been the two main approaches to quantum gravity: the covariant and canonical programs (string theory is not covered). The quantization of gravity is discussed by analogy with the quantization of the electromagnetic field.  The conceptual and technical problems of both approaches are discussed, and the paper concludes with a discussion of evidence for quantum gravity from the rest of physics.

The paper assumes some familiarity with non-relativistic quantum mechanics, special relativity, and the Lagrangian and Hamiltonian formulations of classical mechanics; some experience with classical field theory, quantum electrodynamics and the gauge principle in electromagnetism might be helpful but is not required.  No knowledge of general relativity or of quantum field theory in general is assumed.
\end{abstract}

\section{Introduction}

Physicists --- so the popular account goes --- have reduced all interactions in the world to four fundamental forces: electromagnetism, the weak and strong nuclear interactions, and gravity.  Forces and particles previously thought to be separate have been unified: electricity with magnetism, protons and neutrons (via the quark model) with pions.  There is even a fairly good unification of electromagnetism with the weak interaction, and Grand Unified Theories adding the strong interaction (quantum chromodynamics) to this pair are constantly discussed.  One often gets the impression that we are on the verge of a Theory of Everything, and that combining gravity with the others is just a matter of technical details.

And yet we do not have that unified theory, for a very good reason: we cannot yet construct a theory in which gravity is treated quantum-mechanically, and there is little hope that such a theory is just around the corner.  General relativity is far richer, conceptually and mathematically, than any of the classical field theories on which particle physics is based, and its quantization has eluded us for half a century, throwing up fascinating insights and open questions in the process.

This review is an attempt to explain what makes this quantum theory of gravity so unattainable, and to show what sort of progress has nonetheless been made.  The strategy is as follows: to study, in parallel, electromagnetism and general relativity, and to show why the techniques that have served us so well in the quantization of the former theory should prove so difficult to apply to the latter.  The mathematical complexities of gravity have been almost entirely omitted, and points relating to them are illustrated instead by analogy with the far simpler electromagnetic theory.

The review makes no pretence to be either complete or up-to-date.  As far as the former is concerned, quantum gravity has become a vast subject and any attempt to survey the literature \emph{in toto} would be doomed.  As for the latter, while there have been exciting developments over the last two decades the mathematical complexity that would be required to discuss them would be beyond the scope of this paper.  In particular, the huge and growing field of string theory is barely mentioned, despite its claim to be the most popular current route to a quantum theory of gravity.  This omission is partly because string theory tries to construct a theory of all interactions, going beyond gravity.  Primarily though, it is simply because the author is not qualified to discuss it.

As such, there is little in this review which would be news to a worker in the field in the early 1980s.  However, the questions raised here were seen as relevant then and by-and-large are still seen as relevant today.  

\section{Classical Electromagnetism in Minkowski spacetime}

\subsection{The action principle}

For the purposes of comparison with gravity, it is easiest to begin with the covariant action form of electromagnetism.  Recall that the theory is defined on the space of U(1) connections on spacetime~\cite{nakahara} --- that is, on the space of real vector fields $\gpt_{\mu}$ on spacetime, identifying any two vector fields which differ by a local gauge transformation (of which more later).  An appropriate action for such a theory is
\begin{equation}
S_D = \int_D \mathrm{d}^4\mathrm{x} \;\lgr
\end{equation}
where
\[\lgr = -\frac{1}{4} \gfs_{\mu \nu} \gfs^{\mu \nu},\]
\[\gfs_{\mu \nu } = \partial_{\mu}\gpt_{\nu}- \partial_{\nu}\gpt_{\mu}\]
with Greek indices ranging over spacetime indices 0,1,2,3, and with $D$ any region of spacetime\cite{jackson}.

What this means is that for a classical field configuration within $D$ to be dynamically acceptable, $S_D$ for that configuration must remain unchanged under small variations of the field vanishing outside $D$.  Introducing a small variation 
\begin{equation}
\gpt_{\mu} \rightarrow \gpt_{\mu}+\delta \gpt_{\mu}
\end{equation}
we get
\begin{equation}
0=\delta D_S = - \int_D \mathrm{d}^4\mathrm{x} \; \gfs^{\mu \nu} \partial_{\mu} [\delta \gpt_{\nu}]
\end{equation}
and integrating by parts gives us the familiar Maxwell equations in their covariant form:
\begin{equation}
\partial_{\mu} \gfs^{\mu \nu} = 0.
\end{equation}

As is well known, this action is invariant under the `gauge' variations
\begin{equation}
\delta \gpt_{\mu} = \partial_{\mu} f,
\end{equation}
where $f$ is any real function on spacetime.  Physically speaking, this is the motivation for interpreting $\gpt$ as a \emph{connection} --- that is, as a rule telling us the phase change for a (quantum) test charge when it is moved along a given path.  Classically we can take \gfs\ to be fundamental and ignore \gpt; quantum-mechanically phenomena such as the Aharanov-Bohm effect(\cite{aharanovbohm,chambers}; \cite[pages 365--369]{nakahara}) demonstrate the experimental consequences of \gpt.

The gauge freedom of the vector field \gpt\ is our motivation for viewing the space of possible field configurations not as the full space of vector fields on spacetime, but as the space of fields modulo gauge transformations.  Viewing this space as a space of connections --- that is, as a space of geometric entities --- facilitates the comparison with gravity, but the reader will lose little by ignoring this aspect in the rest of this review.

\subsection{Splitting spacetime into space and time: the 3+1 formalism}
\label{3+1split}

In this section, we break the manifest relativistic covariance of electromagnetism, essentially by choosing a preferred Lorentz reference frame.  This is essential if we are to understand the dynamics of the theory, for dynamics is about the evolution of states through time and cannot be considered without picking out a preferred notion of time.  Furthermore, this split is necessary to go to phase space and the Hamiltonian formalism, which we need to do in order to formulate a quantum theory.

The reader who feels that to make such a break is to do violence to relativity, may be comforted by the following observation.  To specify a system's state, we have to give its position in phase space: that is,  the positions and momenta of all its constituent parts\footnote{In the case of a field theory, this means the state of every point of the field at a given time.}. However, once we know the positions and momenta we may calculate the state of the system at all times, and conversely if we know its state at all times we know where in phase space it is at any given time.  The upshot of all this is that points in phase space are in 1:1 correspondence with solutions to the dynamical equations --- and the space of solutions is not in any way dependent on our choice of reference frame.  This `covariant phase space' formalism will not be used further in this review, but is discussed in \cite{woodhouse} and (in the context of quantum gravity) \cite{carlip}.  Further issues to do with the non-covariance of the theory will be discussed in section \ref{why3+1}.

To make the split is very simple: choose a set of inertial coordinates on spacetime and choose our region $D$ to be all points with time coordinates between $t_1$ and $t_2$.  Then the action is
\begin{equation}
S = \int_{t_1}^{t_2} \mathrm{dt} \int \mathrm{d}^3\mathbf{x} \lgr.
\end{equation}
Breaking \gpt\ into its components $\gpt = (V,\vctr{A})$ we get the unlovely expression~\cite{matschull}
\begin{equation}
S = \int_{t_1}^{t_2} \mathrm{dt} L
\end{equation}
where
\begin{equation}
L= \frac{1}{2}\int \mathrm{d}^3\mathbf{x} \left( \left( \vctr{\nabla} V - \frac{\dr{\vctr{A}}}{\dr{t}}\right) ^2 - \left( \vctr{\nabla}\times \vctr{A} \right)^2 \right).
\end{equation}

Though this may look unfamiliar, it is now in the form we normally associate with nonrelativistic, classical-mechanics system: the dynamics are given by a Lagrangian which is a function of the configuration of the system at a given time and of the rate of change of that configuration.  The only difference is that here, by `configuration' we mean the \vctr{A}\ and $V$ fields on 3-dimensional space at a given time.  To extract the dynamics we can use Hamilton's variational principle
\begin{equation}
0= \delta \int_{t_1}^{t_2} \mathrm{dt} L
\end{equation}
(note that this is just a thinly disguised version of the covariant action principle used above)
from which follows the Euler-Lagrange equations.  In fact these will turn out to be the familiar Maxwell equations, but we shall derive these later from the Hamiltonian viewpoint instead.

Given this format, it is natural to ask what is the `configuration space', which must be analogous to the set of particle positions in classical mechanics.  We might expect it to be the space of pairs $(V,\vctr{A})$ but this is spoiled by gauge invariance.  From elementary electromagnetism we know that $V$ may be adjusted arbitrarily, so we do not treat it as a dynamical variable.  But even the space of vector fields \vctr{A}\ is too large, as we can always add the gradient of an arbitrary function to \vctr{A} without changing the physical state.  This issue will be discussed further in section \ref{emphasespace}.

\subsection{The Hamiltonian formalism}

To quantize a classical theory we usually begin with its description in the Hamiltonian formalism.  To facilitate this we need the Hamiltonian form of electromagnetism.  

Recall that in the case of finitely many point particles with coordinates $q^1, \ldots, q^n$, we define the momenta $p_1, \ldots, p_n$ by~\cite{goldstein}
\begin{equation}\label{legendre}
p_j = \pbp{L}{\dot{q}^j}
\end{equation}
and rewrite Hamilton's principle as

\begin{equation}
\label{hp}
0=\delta S = \int_{t_1}^{t_2} \mathrm{dt} (p_i \dot{q}^i - H(q,p))
\end{equation}
where
\[H(q,p) = p_i \dot{q}^i - L(q,\dot{q})\]
(in which $\dot{q_i}$ is obtained as a function of $p,q$ by inverting (\ref{legendre})).  Then varying the path in Hamilton's principle gives Hamilton's equations
\begin{equation}
\dot{q}^j= \pbp{H}{p_j}
\end{equation}
and 
\begin{equation}
\dot{p_j}= - \pbp{H}{q^j}.
\end{equation}
This procedure will also apply for electromagnetism, with two changes.  The first is that electromagnetism has infinitely many degrees of freedom, with three coordinates per space point.  This means that the momentum will be another vector field, which (with the benefit of hindsight) we will denote as $\vctr{E}$.  The other is due to the gauge freedom, which means that not all the coordinates are true degrees of freedom.
The infinite-dimensional form of (\ref{legendre})\ will be
\begin{equation}
\vctr{E} = \frac{\delta L}{\delta \dot{\vctr{A}}}
\end{equation}
and a quick calculation gives
\begin{equation}
\vctr{E}=\dot{\vctr{A}}-\vctr{\nabla}V ;
\end{equation}
of course this is just the expression for the electric field in terms of the potential, justifying our notation.  The Hamiltonian turns out to be\footnote{To derive this we have had to integrate by parts, which necessitates our assuming the fields to drop off sufficiently quickly at infinity.}~(\cite{matschull}; \cite[pages 450--469]{wald})
\begin{equation}
H = \int \mathrm{d}^3\mathbf{x} \left(\frac{1}{2}\vctr{E}^2 + \frac{1}{2}\vctr{B}^2 + V \vctr{\nabla} \cdot \vctr{E} \right)
\end{equation}
where we have defined $\vctr{B} = \vctr{\nabla} \times \vctr{A}$.
Varying \vctr{E}\ and \vctr{A}\ in (\ref{hp})\ gives analogies of Hamilton's equations:
\begin{equation}
\dot{\vctr{E}} = \vctr{\nabla} \times \vctr{B}
\end{equation}
and
\begin{equation}
\dot{\vctr{A}} = - \vctr{E} - \vctr{\nabla}V
\end{equation}
which on taking the curl of both sides becomes
\begin{equation}
\dot{\vctr{B}} = - \vctr{\nabla} \times \vctr{E}.
\end{equation}
However, to get all of the dynamics we also have to vary the action with respect to $V$, which gives Gauss's Law\footnote{The fourth Maxwell equation, $\vctr{\nabla} \cdot \vctr{B} = 0$, follows trivially from the definition of \vctr{B}.}:
\begin{equation}
\label{gausslaw}
\vctr{\nabla} \cdot \vctr{E} = 0.
\end{equation}
This equation has a fundamentally different structure from the other dynamical equations above: while they tell us how \vctr{E} and \vctr{A} evolve from given initial conditions, Gauss's Law puts a constraint of what those initial conditions \emph{are}.  To be more specific, not all vector fields are admissible momenta: only those which are divergence-free can be used.  An equation like this is called a constraint~\cite{matschull,wald}, and will appear whenever a theory has gauge freedom.

\subsection{Phase Space}
\label{emphasespace}
Notionally, then, the phase space of electromagnetism consists of pairs (\vctr{A},\vctr{E}) of vector fields on 3-space, implying that it has six dimensions per space point.  However, this is complicated by the gauge freedom: in the space of \vctr{A}-fields we should really treat any two fields differing by a gauge transformation as the same field (leaving two dimensions per space point), and in the space of \vctr{E}-fields we have to impose the constraint $\vctr{\nabla}\cdot\vctr{E}=0$ at every space point, again reducing the dimensionality of the space of \vctr{E}-fields from 3 to 2 per space point\footnote{Here we see the reason to expect constraints in the theory: the `position' \vctr{A}\ and the `momentum' \vctr{E}\ should be dynamically conjugate, so if gauge invariance lowers the actual degrees of freedom of \vctr{A}-fields then something else must restrict the degrees of freedom of \vctr{E}-fields.}.

This leaves electromagnetism having two degrees of freedom per space point, which accords with our quantum description in terms of massless spin-1 particles with two orthogonal spin states.  

This restriction of the apparent phase space by gauge invariance and constraints is characteristic of gauge theories, and makes it hard to describe and classify the true dynamical states.  Two normal ways of doing it are
\begin{enumerate}
\item To choose a particular gauge and describe the theory in that gauge, and
\item To find some way of describing the states that is itself gauge-invariant.
\end{enumerate}
In general both of these (and especially the second) can be very difficult. In fact, in the case of electromagnetism the theory is simple enough that we can do either: there exist simple gauge choices like $\vctr{A}^3 = 0$ (axial gauge) or $\vctr{\nabla}\cdot \vctr{A}=0$ (London gauge) which can be imposed at all times and for all solutions, and we can usefully describe the configuration space as 
\begin{quote}
The space of equivalence classes of vector fields differing only by the gradient of a potential
\end{quote}
(note that this is only a useful description because the mathematical description of a gauge transformation in electromagnetism is easy to understand).

\section{Classical General Relativity}

\subsection{The dynamics of spacetime}
\label{dynamics}

Electromagnetism is a theory describing how certain fields evolve on a fixed spacetime background.  So is Dirac field theory; so is QCD; so is every field theory used in modern physics, with one exception.

General Relativity differs from every other field theory in that spacetime itself is the dynamical entity under consideration: `gravitation' occurs because curved space affects the path of matter.  In Wheeler's famous words~\cite{MTW},
\begin{quote}
Space acts on matter, telling it how to move.  In turn, matter reacts back on space, telling it how to curve.
\end{quote}
Unlike in the case of Newtonian gravity, the `gravitational field' (\ie curvature of spacetime) can propagate and have degrees of freedom even without matter --- so vacuum gravity contains interesting objects, such as black holes and gravity waves.

The field equations for general relativity are highly complex, consisting of ten coupled non-linear partial differential equations, and I shall not write them here (see \eg \cite{MTW, wald}).  This makes finding exact solutions extremely difficult and only a few are known, usually corresponding to situations with a high degree of symmetry (such as black holes and homogenous cosmologies).  However, in the limit of a very weak field, we can approximate spacetime as nearly flat and look at the dynamics of the small variations from flatness.  In this limit the equations become linear and can be solved approximately (this is the so-called `weak field' approximation~\cite{MTW, wald}).  

The fact that general relativity does not have a background spacetime makes it difficult to treat conceptually as well.  In electromagnetism, for instance, we are entitled to use the background spacetime for reference: it is meaningful to say `at spacetime point $x$ the value of the \gfs-field was such-and-such'.  This does not work for gravity: to say `at spacetime point $x$ the structure of spacetime was such-and-such' implies that we have a meaningful way of referring to spacetime points in the absence of spacetime, which clearly is nonsense\footnote{Some of the philosophical consequences of this are discussed in \cite{earman}, Chapter 9, and references found therein.}.

\subsection{Gauge invariance in general relativity}

The `gauge' invariance of general relativity actually corresponds to the difficulties noted at the end of the last section, in giving coordinates without reference to spacetime.  However, for our purposes it can more simply be understood as follows.  

To describe the dynamics of gravity, we need --- as with electromagnetism --- to split spacetime into space and time.  However, in a theory where spacetime itself is dynamic there can be no preferred way of making this split.  We have to cut 4-dimensional spacetime up into 3-dimensional slices, but in general we can do this in any way we like, freely choosing curved, bumpy slices.  Any such cutting defines a concept of time, because we move forward in time by moving from one slice to the next.  Gauge transformations correspond to a change in the definition of time, by changing our choice of how to slice.  (This arbitrariness is often referred to as `many-fingered time'; see \cite{MTW} for a discussion restricted to the classical context).

The mechanics of making a split were developed in the work of Arnowitt, Deser and Misner~\cite{adm} on canonical general relativity; see, \egc, \cite{MTW} for a brief treatment, and \cite{isenbergnester} for a full survey.

To understand better how much choice we have in the definition of time, we contrast general relativity with two previous theories of spacetime.
In Newton's conception of spacetime\footnote{Originally published in \cite{principia} and discussed extensively in \cite{earman}.} there is a unique way of making the division: time is universal, and this concept makes instrumental sense since there is no upper limit on speed of signal propagation in Newton's physics\footnote{I am grateful to Harvey Brown for this observation.}.

In special relativity we have more freedom.  As is widely known there is no preferred reference frame in special relativity.  However, there is still a preferred class of reference frames: inertial reference frames, \ie those with Lorentz coordinates, related by Lorentz transformations.  In the language used here, these preferred frames correspond to a preferred family of slices, and this family is in fact the family of \emph{flat} slices.  But the only reason that there exist such flat slices is because, in special relativity, spacetime itself is flat.  If we take a highly non-flat spacetime --- as is the generic case in general relativity --- flat slices simply do not exist.

In fact, the difference between the freedom of choices of slicing in special relativity and in general relativity is analogous to the difference between local and global gauge transformations in quantum mechanics~\cite{kaku,nakahara}.  Given a quantum wave-function, we can always make a global transformation 
\begin{equation}
\psi(x) \rightarrow  \mathrm{e}^{i \, \phi} \psi(x)
\end{equation}
but to make a local transformation
\begin{equation}
\psi(x) \rightarrow \mathrm{e}^{i \, \phi(x)} \psi(x)
\end{equation}
requires us to introduce the connection \gpt\, which in turn inspires us to make this connection dynamical.  Similarly the Lorentz transformations of special relativity are necessarily global, \ie affect the definition of time across all of spacetime, whereas in general relativity we can deform the slicing surfaces in a small region without affecting them outside that region.  Much effort (\eg \cite{gockeler, hehl, hehl2, neeman, trautman}) has been put into making this more than an analogy, and trying to understand general relativity as the gauge theory built around special relativity.  This program has had a degree of success, but unfortunately it has also led to much confusion and has tended to obscure the significant structural and conceptual differences between general relativity and other gauge theories.

In fact, one of these conceptual differences will be very important for us here.  In electromagnetism two configurations differing only by a gauge transformation are regarded as different descriptions of the same physical state.  However, in general relativity one possible `gauge transformation' will be to move all of the slices forward through spacetime.  If we keep our eye on a given slice, that slice will then describe a `later' section of spacetime --- in other words, time evolution is just one form of gauge transformation, and if we identify gauge-equivalent configurations then we will freeze out the time evolution~\cite{carlip,matschull}.  Classically this is by and large a curiosity; in the quantum theory it is a ferocious problem, as we shall see.

\subsection{Phase space}
\label{grphase}

To quantize gravity, we would like a description on phase space, and --- again by analogy with electromagnetism --- this turns out to be nontrivial because of the gauge transformations.  We once again make some splitting of spacetime into space and time; note that the degree of arbitrariness here is much greater than for electromagnetism as we may no longer select a Lorentz slicing\footnote{Of course, the only reason we were able to select such a slicing before is that we were considering electromagnetism in \emph{Minkowski} spacetime.  If we were to study the propagation of electromagnetic fields on some more arbitrary fixed spacetime, this selection would not be possible.}.
The configuration space of the theory, if we ignore the gauge freedom, is then the space of 3-dimensional geometries; the phase space turns out to include, in the guise of momenta, information on how to embed that three-dimensional geometry into a 4-dimensional one.  The Hamiltonian is fearsomely complicated (see \cite{wald} for a derivation, and \cite{fischermarsden} for a detailed discussion) but its derivation follows the same pattern as for electromagnetism --- it just has more algebra.

Of course, we have no business ignoring the gauge freedom.  As with electromagnetism, we will find that some configurations should be identified as being gauge transformations of one another, and that there will be some constraint (the so-called `Hamiltonian' constraint') on the admissible states.  This constraint is extremely complicated, and in general it is a very hard problem to check if any given state satisfies it\footnote{This is a significant oversimplification.  There are other, `momentum', constraints corresponding to changes in the coordinates used on the spatial slice, but these are conceptually and technically much simpler to handle and will be omitted hereafter.}.

So far all of this is in accordance with electromagnetism - but now remember that gauge transformations correspond to changes in the definition of time.  If we are identifying two states as identical under gauge transformations, we are just saying that they describe the same system at different times.  No wonder that the constraint is so hard to solve:  it encodes the dynamics for the entire system!  

In fact, this is why we call the constraint `Hamiltonian'.  It is responsible for the time evolution, and turns out to be equal to the actual Hamiltonian function.  A corollary of this is that --- since by definition the constraint has value zero on admissible states --- all physical states have zero energy\footnote{Actually, this only holds for `compact' spaces, such as a closed universe.  It is possible, by judiciously keeping track of boundary terms, to define energy, momentum \etc for an open universe in which the matter density and spacetime curvature tend to zero with large distances: see \cite{wald} for an introductory discussion, and \cite{ashtekar2,ashtekar1} for a full treatment.}.

As should be apparent from the above, understanding the true degrees of freedom of gravity is fearsomely difficult --- far harder than for electromagnetism.  The second strategy in section \ref{emphasespace} (finding a simple gauge-invariant description) seems all but impossible, as it would amount to solving the field equations in the general case.  As for choosing a specific gauge, this is extremely difficult to do: how can we specify a way of slicing spacetime that will make sense irrespective of the contents of that spacetime?  In fact one way is known~\cite{york1,york2,MTW} (the so-called `York time') but it has proved difficult to establish whether this method really is well defined.  In any case, being restricted to a specific gauge is very inconvenient in general relativity (far more so than in electromagnetism) since we have lost our freedom to define time as we wish --- analogous to, but more serious than, being forced to do all of special relativity in a specific frame. 

\section{Quantizing the electromagnetic field}

Clearly there is something unsatisfactory about the whole notion of quantization.  Presumably, the quantum theory is the more fundamental and we should begin from a quantum theory and then `classicalize' it.  Nonetheless, quantization remains the only program we have to generate realistic quantum field theories, so we are forced to use it.  But it is important to bear in mind that the quantization process is a series of rules-of-thumb rather than a well-defined algorithm, and contains many ambiguities.  In fact, for electromagnetism we shall find that there are (at least) two different approaches to quantization, and that while they appear to give the same theory they may lead us to very different quantum theories of gravity.

Before these methods are described, however, it is necessary to say something about covariance.  The reader who is unconcerned about violations --- real or apparent --- of covariance may skip the next section. 

\subsection{Canonical vs. Path Integral approaches}
\label{why3+1}

Both of the approaches to quantization to be detailed below rely, as a first step, upon casting the classical theory into Hamiltonian form and thus breaking its manifest covariance.  For all that phase space may be regarded in some sense as covariant (see section \ref{3+1split}), and for all that the final quantum theory may have some measure of covariance restored to it, nonetheless it would be nice to have a manifestly covariant way to proceed.

On the surface, there is such a way: the Feynman path-integral approach to quantization~\cite{feynhibb, kaku, weinberg} begins with the Lagrangian, rather than the Hamiltonian, formulation, and calculates all observable quantities in terms of the path integral
\begin{equation}
\int \mathcal{D} \gpt \exp \mathrm{i} \left( \int \mathrm{d}^4\mathrm{x} \;\lgr [\gpt] \right)
\end{equation}
where the second integral is over all of spacetime and the first is over all gauge-inequivalent \gpt-fields; this integral is manifestly covariant.

Unfortunately, appearances can be deceptive.  To extract physical content from this path integral we need rules to go from the vacuum expectation value above to the expectation values for scattering between particles; these rules (the LSZ reduction formulae~\cite{itzyksonzuber, kaku}) and indeed the existence of the particle spectrum itself require the canonical formalism to derive.  They, and indeed the path integral itself, fade into the background once the Feynman rules are derived, but for all that they are required in the foundations.

It may be objected that Feynman's path integral approach is usually described as an alternative formalism for all of quantum mechanics, and not just a calculation tool.  Indeed this is so: the fundamental rule of the path integral approach is that, for two configuration-space eigenstates $\ket{q_1;t_1}$ and $\ket{q_2;t_2}$, we have
\begin{equation}
\bk{q_1;t_1}{q_2;t_2} = 
\int _{q=q_1, t=t_1}^{q=q_2; t=t_2} \mathcal{D} q
\exp \left( \mathrm{i} \int^{t_2}_{t_1} \mathrm{d}t L[q(t)] \right)
\end{equation}
and from this all of the dynamics can be deduced.  The problem is that, for a field theory, a `point' in configuration space is a field configuration at a given time~\cite{weinberg} --- and of course, to define such a thing we have to choose a preferred time.

This example shows why we should not be too surprised to find problems with covariance when we try to quantize a theory.  Ultimately states in quantum mechanics are \emph{non-local}: entanglement makes it clear that a description in terms of isolated spacetime events simply will not do.  But then any attempt to describe nonlocal states will struggle to avoid the problem of choosing preferred reference frames\footnote{This is not to say that this problem cannot be avoided.  The Heisenberg picture, with its time-independent states and operators which vary across spacetime, is much closer to manifest covariance --- but the theory cannot, as far as I know, be formulated in this picture using the path-integral approach.  When we consider the measurement problem, covariance issues become still more fierce: it appears impossible to carry out a collapse of the wave-function without violating at least the spirit of relativity, and one of the major strengths of Everett-style interpretations~\cite{everett,deutsch96} is the promise they hold out for a quantum theory which is more naturally compatible with relativity~\cite{saunders,wallace}.  See \cite{maudlin} for a discussion of the tension between the measurement problem and Lorentz covariance.}.

See \cite{ashtekar3} for a discussion of the relationship between the canonical and path-integral methods, in the context of quantum gravity.
\subsection{Particle quantization}
\label{emparticle}

The most common algorithm for quantizing electromagnetism goes as follows (I follow the treatment of \cite{landauqft}).  Firstly, we assume the system to be confined within periodic boundary conditions (`putting the field in a box'); this step is fairly easily avoided but the presentation is simplified if we leave it in.  A choice of gauge is made (London gauge is convenient for conceptual purposes; a relativistically covariant gauge is often easier for calculations) and the 4-potential \gpt\ is expanded in terms of plane waves:
\begin{equation}
\gpt_{\mu}(\vctr{r},t)= \sum_{\vctr{k},\alpha}\left( c_{\vctr{k},\alpha} \gpt_{\vctr{k},\alpha} + c^*_{\vctr{k},\alpha} \gpt^*_{\vctr{k},\alpha}\right)
\end{equation}
where
\begin{equation}
\gpt_{\vctr{k},\alpha}= \sqrt{4 \pi} 
\frac{\mathrm{e}^{i (\vctr{k}\cdot \vctr{r}-\omega t)}}{\sqrt{2\omega}}
\vctr{e}^{\alpha}.
\end{equation}
Here the $\vctr{e}^{\alpha} (\alpha =1,2)$ are two orthogonal polarisation 4-vectors, whose precise form is given by (and determines) the choice of gauge, and the $c_{\vctr{k},\alpha}$ are expansion coefficients.  Writing the Hamiltonian in this plane-wave expansion we get
\begin{equation}
\label{emham}
H[\gpt] = \sum_{\vctr{k},\alpha}\frac{1}{2}\omega \left(
c_{\vctr{k},\alpha}c^*_{\vctr{k},\alpha} + c^*_{\vctr{k},\alpha}c_{\vctr{k},\alpha}
\right)
\end{equation}
Guided by a mixture of theoretical\footnote{Mostly the similarity of the expression for the Hamiltonian to that of a system of harmonic oscillators.} and experimental reasoning, we now reinterpret the theory as describing a system of spin-1 particles (photons) for which the $c_{\vctr{k},\alpha}$ are annihilation operators.  This consists of promoting the $c$,$c^*$ to the status of operators $c$,$c^{\dagger}$ and imposing the commutation relations
\begin{equation}
\comm{c_{\vctr{k},\alpha}}{c^{\dagger}_{\vctr{k'},\alpha '}} = \delta_{\vctr{k},\vctr{k'}} \delta_{\alpha,\alpha '}.
\end{equation}
The free-particle Hamiltonian now describes a collection of independent, free particles.  To add interactions, we replace terms like $\gpt_{\mu} J^{\mu}$ in the Hamiltonian with their operator equivalents.  Note that this process is fraught with ambiguities because quantum operators do not commute: for instance, the quantum form of the Hamiltonian (\ref{emham}) changes by an infinite constant if we replace $1/2 (c c^* + c^* c)$ by $c^* c$ in the classical version.  Operator-ordering problems like this are a major source of ambiguities in quantization.

We pause a moment to consider what this theory is describing: a universe made up of particles moving through spacetime.  For electromagnetism in the absence of sources, this quantization can be treated as exact; for most other quantum field theories there is self-interaction and we have to treat the self-interaction terms as perturbative corrections.  Doing dynamics in this framework is mostly restricted to considering scattering situations: a collection of free particles far away from one another collide and in due course some other collection of free particles are observed.  Some non-perturbative calculations can be done, but the lack of general methods for non-perturbative quantum field theory remains a major lacuna of the theory (especially in quantum chromodynamics where it is required to explain phenomena like quark confinement and spontaneously broken axial symmetry~\cite{chengli,shuryak}).  The formalism we have constructed here would seem to have difficulties even expressing such non-perturbative ideas, since its construction is perturbative in nature.

To do perturbative calculations, we have an algorithm \emph{par excellence}: the Feynman diagrams, which can quickly be derived from the quantization procedure above~\cite{weinberg} and lend themselves naturally to an interpretation in terms of scattering events.  Further, the Feynman diagrams are manifestly covariant, even if the formalism we used to derive them was not.

Actually doing calculations with Feynman diagrams rapidly reveals the presence of infinitely large terms, which have to be removed by sleight-of-hand: this `renormalisation' process is now mathematically fairly well understood even if it is still physically somewhat mysterious~\cite{binney,kaku}.  Essentially whether or not a theory is renormalisable depends on the number of lines coming into any vertex in a Feynman diagram: if there are more than a certain number of lines then the number of infinities that have to be removed will increase with the complexity of the diagram, and in removing an infinite number of infinities all predictive power would be lost.  Fortunately this does not occur with electromagnetism: each vertex on a Feynman diagram corresponds to an interaction term (such as $\gpt_{\mu}\psi^{\dagger}\gamma^{\mu} \psi$ in QED) and the number of lines entering the vertex is equal to the power of the fields in the interaction term (so, for instance, in the term above there is one line corresponding to the \gpt -field and two corresponding to the $\psi$-field, for a total of three.  This is the maximum permissible for a theory of electrons and photons, but it \emph{is} permissible: QED is renormalizable~\cite{kaku}).  These renormalisation considerations show why we would have difficulty with a theory with non-polynomial interaction terms: to write down the Feynman rules for such a theory we would need to make a power-series expansion, which would contain terms of all powers.  Most of these terms would be nonrenormalisable.

\subsection{Field quantization}
\label{emfield}

While the method of the previous section seems to correspond nicely to what we observe in particle-physics experiments, it is at least noteworthy that it seems to bear little resemblance to the quantization procedure used for ordinary mechanical systems (such as a point particle).  

Recall that in this case, the quantum system is described by a wave-function on the configuration space, \ie a function $\psi$ from the configuration space to the complex numbers.  The physical interpretation of this wave function is that, if $U$ is some subregion of configuration space, the probability of finding the system to be in $U$ on measurement is
\begin{equation}
\Pr (\mbox{system in }U) = \int_U \dr{\vctr{x}} | \psi(\vctr{x}) |^2.
\end{equation}
The position functions $q^1, q^2, \ldots, q^n$ are replaced by multiplication operators:
\begin{equation}
( \widehat{q^j} \psi ) (\vctr{x}) = x^j \psi (\vctr{x})
\end{equation}
and the momentum functions $p_1, \ldots, p_n$ by derivative operators:
\begin{equation}
\widehat{p_j} \psi = - \mathrm{i} \pbp{\psi}{x^j}.
\end{equation} 
Functions of $p$ and $q$ are defined similarly, although major problems \emph{may} result from operator-ordering ambiguities: classically, $pq=qp$, but this does not hold at the operator level.

In field quantization, we try to duplicate this process for the electromagnetic field (here I follow the treatment of \cite{matschull}).  The analogies we need are (ignoring, for the moment, subtleties due to gauge invariance):

\vspace{7mm}
\begin{tabular}{ll}
Configuration space & Space of vector fields \vctr{A} \\ & \\
Position functions $q^i$ & Potential functions $\vctr{A}^i(x)$ \\
& at each point x \\ & \\
Momentum functions $p_i$ & Field-strength functions $\vctr{E}_i(x)$ \\
& at each point x \\
\end{tabular}
\vspace{7mm}

So our quantum description of a state will be some sort of wave-functional $\Psi[\vctr{A}]$ which assigns a complex number to each vector field.  The physical interpretation of this wave functional will be that, if $U$ is a collection of \vctr{A}-fields, the chance of the system having vector potential in $U$ upon measurement will be
\begin{equation}
\Pr (\mbox{system in }U) = \int_U \mathcal{D} \vctr{A} | \Psi[\vctr{A}] |^2
\end{equation}
(albeit this equation is largely figurative because of the difficulties in making rigorous sense of functional integrals\footnote{Actually, for a gauge theory there is another problem: this integral, as written, integrates over all \vctr{A}-fields and not just the gauge-inequivalent ones.  Strictly speaking we should introduce some sort of Fadeev-Popov gauge-fixing term into the functional integral~\cite{chengli}, but this will not have any particularly interesting consequences.}).  The `position' operator will be 
\begin{equation}(\widehat{\vctr{A}^i(x)} \Psi ) [\vctr{A}] = \vctr{A}^i(x) \Psi[\vctr{A}]
\end{equation}
and the `momentum' operator will be
\begin{equation} \widehat{\vctr{E}_i(x)} \Psi = - \mathrm{i} \frac{\delta \Psi}{\delta \vctr{A}^i(x)}.
\end{equation}
The Hamiltonian becomes
\begin{equation}
\hat{H}= \int \dr{^3 x} \frac{1}{2}\left( \widehat{\vctr{E}(x)} ^2 + \left( \nabla \times \widehat{\vctr{A}(x)} \right)^2 \right) + V \nabla \cdot \widehat{\vctr{E}(x)}
\end{equation}
(note that $V$ is not quantized, since it is not genuinely dynamical but just chooses the gauge).  Factor-ordering problems will not occur here because the $\widehat{\vctr{A}}$ and $\widehat{\vctr{E}}$ operators are not mixed.

There is one important difference between this system and the ordinary mechanical systems: the constraint equation (\ref{gausslaw}).  It turns out~\cite{kaku} that we cannot impose it as an operator equation because this will be too restrictive (it would be analogous to requiring the classical constraint equation to hold for all vector fields, not just the dynamically acceptable ones!)  Instead we impose it as a condition on states:
\begin{equation}
\nabla \cdot \widehat{\vctr{E}(x)} \ket{\psi} = 0.
\end{equation}
Clearly this will restrict us to a subspace of the original Hilbert space: the space of physical states.

In fact, it turns out that we have a bonus: the physical states are already gauge invariant.  This is because, as it turns out, the constraint is the infinitesimal generator of gauge transformations!  As such, any state which it annihilates will be gauge invariant (the technical details may be found in the appendix).  This is a completely general phenomenon, holding for any gauge theory.

For electromagnetism, we can reach this directly: recall that the `true' configuration space of the system is the space of equivalence classes of vector fields under gauge transformations.  We could have avoided any mention of the constraint by starting off with wave-functionals on \emph{this} space, \ie with functionals $\Psi$ for which
\begin{equation}
\Psi [\vctr{A}+\nabla f] = \Psi [\vctr{A}].
\end{equation}
Effectively, we have imposed the constraint classically and only then quantized.
However, the only reason that this trick works is that electromagnetism  is simple enough to allow this direct description of the reduced configuration space (see \cite[pages 450--469]{wald}).  In general --- and for gravity in particular, as we shall see --- we must impose the constraint at the quantum level.

To summarise: physical states are those annihilated by the constraint operator, and these states are gauge-invariant.

The duality between particle and field descriptions of a quantum field theory means that these two methods of quantization produce physically equivalent theories, differing only by a unitary transformation.  However, when we attempt to quantize gravity it is not clear that the duality will continue to hold: indeed, the two main approaches to quantum gravity follow the two different methods, and seem to have very little in common!

\section{Covariant quantum gravity}

\subsection{Parting of the ways}
\label{parting}

Quantum gravity is a field driven almost entirely by theoretical considerations, with a crippling shortage of experimental data.  As such, it has --- for decades --- been split into different schools, following different methods to construct different theories.  Only when (and if) any one of these schools produces a theory developed enough to make experimental predictions, will it be possible finally to adjudicate between them\footnote{Of course, it would be nice to suppose that the eventual theory would be some sort of synthesis of the various approaches --- but currently we have no way of knowing how likely this is.}.  

The two main approaches to quantum gravity --- ironically they both go back to the same set of papers by Bryce Dewitt~\cite{dewitt1, dewitt2, dewitt3} --- are known in the literature as `covariant' and `canonical' quantum gravity. An informal survey of published papers in 1998~\cite{ishamtalk} estimated that about 70\%\ of publications were on covariant gravity (or more accurately its descendants, primarily string theory) and 30\%\ were on the canonical theory.  In addition, a small minority of papers covered a myriad other approaches: the Euclidean path integral approach of Hawking and co-workers~\cite{hawkingepi}; the twistor theory pioneered by Penrose~\cite{penrosetwistorbook, penrosetwistorpaper}, \etc, none of which will be discussed here.

\subsection{Formulating covariant quantum gravity}

The idea of the covariant approach to quantum gravity is to develop the theory along the lines seen in particle physics: that is, through interacting particles, Feynman diagrams and the like.  (The use of Feynman diagrams is the justification for calling the theory `covariant'.)

To do this, we need to follow as far as is possible the particle quantization method of section \ref{emparticle}, so we need an expansion of a general solution to the general-relativistic field equations in terms of plane waves.  In general this will be impossible because of the extreme nonlinearity of general relativity, so we begin with the weak-field approximation mentioned in section \ref{dynamics} and treat the nonlinear terms as perturbations~\cite{kaku}.  The weak-field approximation \emph{is} linear, and we can make a plane-wave expansion; quantization reveals that the particle interpretation of gravity is in terms of massless spin-2 bosons: `gravitons'.  We now include interactions, including the self-interactions due to the nonlinearity of general relativity, using Feynman diagrams.

\subsection{Conceptual problems}
\label{covconcept}

It should be clear that this approach is essentially perturbative: we have taken an intrinsically non-linear theory, treated it as linear, and added its nonlinear part as a correction.  Now this is not automatically a mistake: it is essentially the approach used in the rest of quantum field theory, after all.  However, it means that the framework is likely to have difficulty accommodating the sort of intrinsically non-perturbative behaviour which we would expect in the vicinity of strong fields.  In fact as was mentioned in section \ref{emparticle} this already causes problems in QFT: specifically in quantum chromodynamics it makes it very difficult to understand features like quark confinement.  In gravity, we might expect similar difficulties in coming to grips with features such as black holes, and other regions where, classically, spacetime is highly non-flat.

And this touches on a more profound problem with covariant quantization.  The theory we have produced is one which describes spin-2 particles propagating through spacetime.  But surely the theory is supposed to be a theory \emph{of} spacetime?  Or put another way, if quantized spacetime `is' a collection of spin-2 particles, in what space do those particles exist\footnote{It might be objected that classical general relativity does not \emph{have} to describe a dynamical spacetime: that the evidence is perfectly compatible with a flat-space field theory that just \emph{looks} as if it describes a dynamical spacetime.  This is of course perfectly true, just as astronomical evidence is perfectly compatible with a geocentric theory that just \emph{looks} as if it describes a solar system with the sun at the centre.  See \cite{deutsch96} for further discussion of this point.}?

The concrete problems that this causes become more apparent when we consider how this quantum framework could describe certain classical features of gravity.  In particular, general relativity, in making spacetime dynamical, allows the causal structure of spacetime to be influenced by entities like gravity waves (the causal structure of spacetime is the structure that defines concepts of `past' and `future'; it tells us which events may causally affect other events, and is the generalisation to curved spacetime of the light-cones of special relativity).  But, as has been stressed by Penrose among others~\cite{joyc}, in covariant quantization the gravitons live in Minkowski spacetime.  The causal structure of this spacetime is \emph{fixed} and no amount of graviton propagation will change it. 

If canonical quantum gravity were to be empirically useful, no doubt these problems could be sidestepped and left to be understood later: this is a method which has worked very well for quantum mechanics in the past.  Unfortunately --- as the next section shows --- this is not the case due to formidable technical difficulties.

\subsection{Technical problems, and further developments}

(This treatment follows \cite{kaku}).

The technical problems in this approach to quantum gravity can be summed up in six words:
\begin{quote}Covariantly quantized gravity is not renormalisable
\end{quote}
It almost goes without saying that this is a disaster.  Nonrenormalisability means that there is no way of proving the infinities in the Feynman diagrams to be controllable: that, on the contrary, that there is every reason to expect them to increase without limit as the Feynman diagrams become more complicated.  Obviously, this makes it completely impossible to extract information from the theory.

To understand why nonrenormalisability should occur, recall the remarks made at the end of section \ref{emparticle}.  There, it was explained that for a theory to be renormalisable the number of lines entering each Feynman-diagram vertex --- equal to the power of the corresponding term in the interaction Hamiltonian --- had to be below a certain finite number.  But the interaction term in general relativity is \emph{non-polynomial} in the fields, so (also as explained in section \ref{emparticle}) we have to make a power-series expansion, leading to an infinite sum of interaction terms each of a higher power than the last.  Clearly, for all but a finite number of these terms renormalisability is impossible.

Even when a theory is formally `nonrenormalisable' in this way there is always the possibility that, through a series of miracles, the various infinities will cancel out.  The formidable mathematical difficulties involved in quantum gravity have made calculations to test this very difficult, but it is now generally recognised that no such fortuitous cancellation is likely to occur above the one-loop level.  (Having said this, it is as well to be cautious: for many years it was thought that \emph{3-dimensional} quantum gravity was nonrenormalisable; then it was shown to be exactly quantizable~(\cite{witten}; discussed in~\cite{carlip}) --- and of course later analysis showed it to be renormalisable all along~\cite{pullin}.)

Most workers in this branch of quantum gravity have so far concentrated on solving these technical difficulties, rather than conceptual ones.  The usual approach has been to modify the classical theory in the hope of producing one with more tractable behaviour.  In the 1970s gravity was coupled with a spin-3/2 field and a new symmetry (supersymmetry) introduced: this was shown to cause `fortuitous' cancellation at the 2-loop level but unfortunately remained divergent at the 3-loop level. Since the 1980s most workers in the field --- driven partly by input from particle physics --- have moved towards string theory, whose Feynman diagrams appear to be not just renormalisable but finite, and which appears to have general relativity as a low-energy limit.  String theory lies beyond the scope of this review and so I shall not discuss it, except to say that (despite recent attention to the issue, \eg \cite{smolinstring}) there is as far as I know no consensus on how to solve the problems of section \ref{covconcept} within a string-theoretic framework.

\section{Canonical Quantum Gravity}

\subsection{Formulating canonical quantum gravity}

In the canonical quantization of gravity, we try to emulate the `field quantization' of electromagnetism discussed in section \ref{emfield}.  To be precise, we start with the phase space of general relativity --- \ie the space of 3-dimensional geometries --- and take our quantum states to be complex-valued wave-functionals on this space (so each wave-functional associates an amplitude to each geometry).  We then impose the Hamiltonian constraint of section \ref{grphase} as an operator equation on the states: this is the famous Wheeler-DeWitt equation~(\cite{dewitt1}; discussed in~\cite{carlip,pullin}).

The technical problems raised here are extremely formidable.  The Wheeler-DeWitt equation is extremely difficult to treat mathematically: indeed the mathematical pathologies here may be worse even than in the covariant case.  In fact, until very recently no solutions at all were known.

In an attempt to get around this, we might try quantizing on the reduced phase space rather than the full one (recall that the reduced phase space consists only of those states satisfying the constraints, and furthermore identifies states differing by a gauge transformation.  In this case the Wheeler-DeWitt equation would automatically be satisfied (essentially we are choosing to solve the Hamiltonian constraint before quantization rather than after it.)  The problem is that explicitly describing the phase space is a virtually hopeless task, equivalent to a general solution of the field equations\footnote{To see why this is, note that to describe the reduced phase space we have to identify states differing by a gauge transformation.  Since in general relativity the dynamical evolution is just a gauge transformation (recall section \ref{grphase}) this would require saying, for any two states, whether one is the time-evolved version of the other --- \ie solving the dynamics.}.

As well as these technical problems, there are formidable conceptual problems.  One of the most acute is the problem of defining the observables of the theory.  Remember that an observable must be gauge-invariant: but with the gauge group of gravity including time evolution, this means that all of the observables must be conserved.  This is a strong restriction, and in fact no-one has yet found any observables at all.  Some intuition as to why can be got when we consider what we mean by observables in (say) electromagnetism: the paradigm observables are the value of the fields at \emph{points in spacetime}.  But, as was stressed in section \ref{dynamics}, when there is no background spacetime to serve as a source of reference points this option is not available to us.  Even if we could find observables it is unclear how they could really describe the universe if they must be time-independent.  This issue is discussed for toy models in \cite{matschull} and for quantum gravity in two space dimensions in \cite{carlip}; the conclusion seems to be that, again, we would need to solve the classical dynamics in all generality to find all of the observables.

\subsection{The Problem of Time}

Perhaps the most profound conceptual problem of the canonical approach --- almost certainly the most extensively discussed --- goes by the name of the problem of time.  The nature of the problem is very simple:
\begin{enumerate}
\item When we quantize a field theory with gauge transformations, the physical states will turn out to be gauge-invariant.
\item In general relativity, time evolution is just a particular gauge transformation.
\item Therefore, physical states of quantum gravity are time-independent.
\end{enumerate}
This can be seen in various other ways: the Hamiltonian of quantum gravity vanishes on physical states; the Wheeler-DeWitt equation contains no mention of time, etc.  Clearly the problem is deeply intertwined with the problem of defining observables mentioned above, but it is simpler and starker: \emph{what has happened to time?}  It is hard to take seriously a theory which denies that systems change in time!

Nonetheless, various strategies for dealing with the problem have been developed.  The general principle is to assume that time evolution is somehow encoded in the other dynamic variables.  This was spelled out --- in the context of ordinary quantum mechanics --- by Page and Wootters~\cite{pagewootters}, and goes as follows.

Suppose the subsystem of the universe in which we are interested has Hilbert space $\hilbert{H}_s$.  Then to actually measure the time-evolution of this system we will need a real, physical clock.  Let us suppose that this clock has Hilbert space $\hilbert{H}_c$.  The clock will also have some set of eigenstates which might be labelled \ket{\mbox{4 o'clock}},\ket{\mbox{5 o'clock}}, etc., and the clock Hamiltonian will carry it from one such state to the next.

So, suppose the system starts in some product state
\begin{equation}
\tpk{\mbox{system at 4 o'clock}}{\mbox{4 o'clock}}.
\end{equation}
Then time evolution will duly carry it, after an hour, into the state
\begin{equation}
\tpk{\mbox{system at 5 o'clock}}{\mbox{5 o'clock}},
\end{equation}
and thereafter into
\begin{equation}
\tpk{\mbox{system at 6 o'clock}}{\mbox{6 o'clock}},
\end{equation}
\etc
But now suppose that we prepare the system (skating blithely over some formidable mathematical problems, including normalisation) in an entangled state,
\begin{equation}
\ket{\psi} = \sum_{\mbox{all times }n} \tpk{\mbox{system at }n \mbox{ o'clock}}{n \mbox{ o'clock}}.
\end{equation}
Then time evolution will carry each element of this superposition into the next, leaving the overall state unchanged.  There is a strong analogy here with plane waves in ordinary quantum mechanics: a plane wave is an equally weighted superposition of all position eigenstates, and it is invariant (up to a phase factor) under any translation.  So this entangled state is an equally weighted superposition of all instants of time and is invariant under time translation.  Nonetheless --- and here is the essence of the Page-Wootters strategy --- the information about time evolution is still present, encoded in the entanglement correlations between the system and the clock.

It is unclear whether this strategy can deal with the problem of time in quantum gravity.  A sceptical note was sounded by Unruh and Wald~\cite{unruhwald}, who (among other objections) pointed out that the notion of time here will at best be approximate.  This is because of a theorem they proved (the `no-clock theorem') to the effect that any realistic quantum clock will have a nonzero probability of running backwards.  Nonetheless, the strategy has been widely discussed: see~\cite{barbour99} and chapter 11 of \cite{deutsch96} for a non-technical discussion, and \cite{barbour1, barbour2} for a detailed and philosophically-inclined analysis.  The implications for observables are discussed by \cite{rovelli}, using the terminology `evolving constants of motion'.

\subsection{Recent progress: the Ashtekar formulation and knot theory}

For most of the 1970s and 1980s relatively little work was done on the canonical approach to quantum gravity: researchers found the Hamiltonian constraint intractable and disliked the non-covariance that seemed inherent in a Hamiltonian description.  This latter problem began to be seen to decline in importance when it was recognised that phase-space can be seen as covariant (see section \ref{3+1split}), but the main impetus for research in the field since the late 1980s was the discovery by Abhay Ashtekar~\cite{ashtekarbook} of a new formulation of classical General Relativity in which the theory developed very strong analogies with other gauge theories, and in which the constraints became much simpler to solve.  Space does not permit a discussion of this formalism (an extremely readable review paper is \cite{pullin}) but it is worth mentioning one discovery which has come out of this work: a strong connection between general relativity and the mathematical theory of knots.

Knot theory, a branch of pure mathematics, is concerned with attempting to find rules to establish when one knot is the same as another (\iec, one can be continuously deformed into another without untying it).  The usual mathematical strategy for problems like this is to find `knot invariants', mathematical objects which can be associated to a knot and which will not change no matter how much the knot is deformed.  With hindsight, such invariants make excellent candidates for gauge-invariant states in quantum gravity: after all, if the quantity is unchanged by being deformed continuously it will be unaffected by gauge transformations.  At time of writing, this topic has yet to be fully explored, but there appear to be deep connections (and the history of modern theoretical physics is filled with examples of deep and useful connections between branches of physics and of pure mathematics: consider Hilbert space and quantum mechanics, or fibre bundles and gauge theory, or vector calculus and Maxwell's equations \ldots).  In particular, knot invariants have indeed furnished us with (a few) genuine gauge-invariant states~\cite{gambini}, and as far as I know are the only such states known. 

In the last decade canonical quantum gravity has been enjoying a renaissance: the technical problems are far from solved but it does at least seem possible to make progress on them, and although the conceptual problems are as fierce as ever it is perhaps an advantage of the approach that it brings them so clearly into view.

\section{Evidence --- experimental, observational, theoretical}

\subsection{Introduction}

As was mentioned in section \ref{parting}, quantum gravity suffers from a crippling lack of experimental findings to give us any indication about the ultimate theory.  We know some such theory must exist to resolve the incompatibility between quantum mechanics and general relativity, but beyond that we are guided mainly by theoretical extrapolation.  In this sense the development of quantum gravity resembles that of classical general relativity, driven primarily by theoretical and aesthetic considerations, more than it does that of quantum mechanics.  Of course --- as with general relativity --- once we have the theory it will doubtless make many further testable predictions.

Furthermore, we are not entirely without guidelines: some of the main ones are sketched below.  Some (few) refer to experimental results; more are due to cosmological observations, or are hints from our existing theories.

\subsection{The parameters of the standard model}

In general, it is hard to get direct data on quantum gravity from experimental particle physics: even the classical gravitational field between two subatomic particles is virtually unobservable, let alone quantum corrections.  Further, the renormalisation procedure which makes it possible to do quantum field theory in the absence of an understanding of deeper theories, also tends to screen out observable effects due to those theories~\cite{binney,kaku}.  This is largely why the Standard Model of particle physics has remained static for twenty years: higher-energy theories tend not to affect the relatively low-energy phenomena which we are able to observe.

Nonetheless the standard model has major unexplained features: primarily the large number (about 20) of physical parameters (masses of particles, fine structure constant, etc.) which have to be inputted by hand~\cite{chengli}.  An explanation of the values of these constants certainly has to lie beyond the Standard Model, and it is at least possible that quantum gravity may have a hand in providing them (dealing as it does with extremely high-energy phenomena).

Further, the explanation given may be of a novel kind.  As is now widely recognised, the values of the physical constants lie within an extremely narrow band of values compatible with the evolution of life.  Small changes in, say, the mass of the electron or the fine structure constant would prevent star formation, or the formation of complex molecules, or any number of other finely balanced procedures.  This implies that the physical constants will not simply be calculable from a higher-level theory, but will somehow have to be selected for --- otherwise the compatibility of the universe with life will be an implausibly large coincidence.  (See \cite{barrowtipler} for the definitive account of these \emph{Anthropic arguments}, and \cite{leslie} and \cite{smolinpaper,smolinbook} for discussions from two very different viewpoints.)   

\subsection{Unsolved mysteries of cosmology}

There are a number of unsolved problems in cosmology which are briefly mentioned here (all are more thoroughly discussed in \cite{peebles}).  None have generally accepted explanations within the current framework of physics, so some information might be gained about any of them by a better understanding of the unknown quantum-gravitational physics in the extreme vicinity of the Big Bang.

\begin{description}
\item[The horizon problem]  The universe, as we observe it, appears strikingly homogenous, with this homogeneity increasing as we look further back in time.  (Fractional fluctuations in the microwave background radiation are only about one part in $10^4$.)  This is the case despite the fact that parts of the current universe far away from one another would have been out of causal contact with one another in the early universe, and so would have had no way to reach equilibrium with one another.
\item[The flatness problem]  Generically, cosmological solutions of Einstein's equations can be open (describing infinite universes, with negatively curved space, expanding forever) or closed (describing finite universes, with positively curved space, doomed to collapse again in finite time).  Only a very precise choice of the density of the early universe would produce a virtually flat universe, right on the boundary between open and closed --- and yet our universe is virtually flat.  To be more precise \cite{peebles} the density of the universe is currently within one part in $10^3$ of the critical density for flatness.  Since the expansion of the universe tends to increase deviations from the critical value, this implies that at the point when our current theories fail us (about $10^{-35}$ seconds after the Big Bang) the density had to have been within one part in $10^{57}$ of the critical value\footnote{It may or may not be a mere curiosity that a virtually flat universe is another requirement for the existence of life (see~\cite{barrowtipler}).}.
\item[The preponderance of matter over antimatter] The current universe contains an overwhelming predominance of matter, but physical laws are virtually symmetric in their treatment of matter and antimatter.  Explanations for this imbalance within the domain of current physics are at best tentative.
\item[Cosmological structure formation]  Current cold dark matter simulations of the expansion of the universe \cite{efstath} produce a good fit to the currently observed large-scale structure, but require as an input a small level of inhomogeneities in the very early universe.  These inhomogeneities have been observed in the microwave background radiation by the COBE satellite, but are still not understood theoretically.
\end{description}
In fairness I should add that there is a possible non-quantum-gravity explanation for the first two of these: the inflation mechanism (discussed in \cite{peebles}).  However, this explanation is still controversial, and in any case needs extremely fine tuning of the cosmological constant (see below!)

\subsection{The cosmological constant}

General relativity contains a free parameter: the cosmological constant, which simulates a constant distribution of perfect fluid across spacetime.  Einstein called its introduction `the biggest blunder of my life'~\cite{gamow} and to this day it is not known empirically whether it is needed in the classical theory: certainly it is empirically indistinguishable from zero.

The situation is complicated by quantum theory.  Quantum field theory assigns an energy to the vacuum of order $\hbar$ per field mode --- even imposing a cutoff around the quantum-gravity scale to make this finite gives an incredible energy density, of order $10^{107}$ joules per cubic metre.  In nongravitational physics this energy has few effects, but it should give rise to an enormous effective cosmological constant, dwarfing the contribution from ordinary matter and totally contradicting observation.  

This is a real dilemma in current physics (albeit one we can usually ignore) and is one of the most explicit clashes between general relativity and quantum theory.  Simply removing the vacuum energy by fiat will not do: the Casimir effect (\cite{casimir}, discussed in \cite{itzyksonzuber}) proves its reality.  The most conservative solution is to postulate a large, negative `real' cosmological constant which cancels the vacuum contribution.  However, this cosmological constant would have to be fine-tuned\footnote{This fine-tuning is again required for the universe to support life.} to better than one part in $10^{120}$.

A recent review of the cosmological constant is \cite{carrolletal}.

\subsection{Singularities}

Even on its own terms (without the challenge from quantum theory) there is ground to think that general relativity needs modification.  This is because the theory frequently predicts the occurence of `singularities' , \ie breakdowns of the predictability of the equations: the big bang is one example, and any black hole contains another.  It is unclear to what extent these singularities require the replacement of general relativity (see \cite{earman2} for a discussion) but the general consensus is that the very high curvatures of space found around them means that quantum gravity must be applied.  Wheeler~\cite{MTW} in particular has strongly advocated the view that the singularities in gravitational collapse are our strongest clue to the nature of quantum gravity.  This view has been championed more recently by Penrose~\cite{penrosesing,penroseenm}, in connection with the arrow of time: the singularity structure of spacetime is strongly time-asymmetric and Penrose has argued that this must be explained within the framework of quantum gravity.

\subsection{Black hole thermodynamics}

Possibly the strongest clue we have to the nature of quantum gravity is the discovery by Hawking~\cite{hawkingradiation} that black holes, contrary to the predictions of classical physics, must emit black-body radiation.  This work has led to some extremely suggestive analogies between thermodynamics and statistical mechanics on the one hand, and gravity on the other.  Since Hawking's original discovery the result has been rederived within the framework of many approaches to quantum gravity, seeming to suggest it is not a mathematical anomaly; see \cite{waldqft} for a review.

Hawking's calculations were, of course, carried out within the framework of a semi-classical approximation, not true quantum gravity; nonetheless there is good reason to believe that the true theory must reproduce his results to be compatible with its classical limit.  There is a nice analogy with the analysis by Einstein~\cite{pais} of the $A$ and $B$ coefficients in atomic transitions: these are derived without using any detailed knowledge of quantum electrodynamics, but nonetheless QED has to conform to them on pain of violating fundamental principles of physics.

\section{Conclusion}

The problems involved in formulating a quantum theory of gravity are ferocious.  Technical and conceptual problems beset us on all sides and there is precious little information from experimental data (one paper which the author has seen claims that quantum gravity will change the spectrum of Hawking radiation slightly --- and then claims that this will be an `experimental test' of quantum gravity!)  

Still, progress has been made, and is continuing to be made.  In a sense, the expectations we have of the completed theory are grand enough that it has to be difficult: different authors seek a resolution of the renormalisation problem in quantum field theory, a taming of the singularities of general relativity, a solution to the measurement (Sch\"{o}dinger-cat) problem, a reason for the direction of time, the unification of fundamental forces, \ldots.  It is far from clear which, if any of these will actually result, but certainly the complete theory will bring rich rewards for our understanding.

Currently, though, the complete theory is not at hand and there is no real reason to expect a breakthrough in the immediate future.  In the words of one worker in the field~\cite{smolinbook},
\begin{quote}We have made great progress, but the fact that general relativity and the quantum are not yet united means that we have no single picture of what the world is that we can believe in.  When a child asks, What is the world, we literally have nothing to tell her. \end{quote}

\section*{Acknowledgements}

I am grateful to Julian Barbour, Harvey Brown, Joy Christian, David Deutsch and Oliver Pooley for discussions of quantum gravity.  

This work is based on a talk given in the Atomic and Laser Physics Department, Oxford University, on June 1st, 1999, and was completed in partial fulfilment of the requirements for transfer from Probationary Research Student to DPhil student at Oxford University. 

\section*{Appendix: The Gauss-law constraint generates gauge transformations}

Recall that, in the field-quantization framework of section \ref{emfield}, the states are wave-functionals $\Psi[\vctr{A}]$ on the space of \vctr{A}-fields and the electric field operator $\widehat{\vctr{E}(x)}$ is defined by 
\begin{equation} \widehat{\vctr{E}_i(x)} \Psi = - \mathrm{i} \frac{\delta \Psi}{\delta \vctr{A}^i(x)}.
\end{equation}

Now, if we had a \emph{finite}-dimensional symmetry group, such as spin, there would be finitely many generators $\hat{\sigma_1}, \ldots, \hat{\sigma_n}$ and a symmetry transformation would be given by
\begin{equation}
\label{symtrans}
\ket{\psi} \rightarrow \exp \left(\mathrm{i} \sum_i \lambda^i \hat{\sigma_i} \right) \ket{\psi}
\end{equation}
where the $\lambda^i$ give the specific gauge transformation (in the case of spin, for instance, they are related to the Euler angles).  The infinitesimal version of (\ref{symtrans}) is of course
\begin{equation}
\label{infsymtrans}
\ket{\psi} \rightarrow \ket{\psi} + \mathrm{i} \epsilon \sum_i \lambda^i \hat{\sigma_i} \ket{\psi}.
\end{equation}

In the case of gauge transformations the gauge group is infinite-dimensional so we replace the $\lambda^i$ with a continuous function $f$ and supply a continuous infinity of generators $\nabla^i \widehat{\vctr{E}_i(x)}$. The analogue of (\ref{infsymtrans}) is then
\[
\ket{\psi} \rightarrow \ket{\psi} + 
\mathrm{i} \epsilon \left( \int \dr{^3 x}  f \nabla^i \widehat{\vctr{E}_i(x)} \right) \ket{\psi}, \mbox{\ie}
\]
\begin{equation}
\Psi[\vctr{A}] \rightarrow \Psi[\vctr{A}] - \epsilon
\int \dr{^3 x}  f \nabla^i 
\frac{\delta \Psi}{\delta \vctr{A}^i(x)}.
\end{equation}
Integrating by parts, this gives us
\begin{equation}
\Psi[\vctr{A}] \rightarrow \Psi[\vctr{A}] + \epsilon
\int \dr{^3 x}  (\nabla^i f)   
\frac{\delta \Psi}{\delta \vctr{A}^i(x)}.
\end{equation}
But this is just the first two terms of an infinite-dimensional Taylor expansion, and we have
\begin{equation}
\Psi[\vctr{A}] \rightarrow \Psi[\vctr{A} + \epsilon \vctr{\nabla} f]
\end{equation}
which is a gauge transformation.  We have proved what we set out to establish: that the constraint operators are the generators of the gauge group.  It follows that any physical state, \ie any state satisfying
\begin{equation}
\nabla^i \widehat{\vctr{E}_i(x)} \ket{\psi} = 0,
\end{equation}
must be gauge invariant.

\bibliography{qg}
\bibliographystyle{abbrv}

\end{document}